\documentstyle[11pt,newpasp,twoside,psfig]{article}
\def\spose#1{\hbox to 0pt{#1\hss}}
\def\lta{\mathrel{\spose{\lower 3pt\hbox{$\mathchar"218$}}
     \raise 2.0pt\hbox{$\mathchar"13C$}}}
\def\gta{\mathrel{\spose{\lower 3pt\hbox{$\mathchar"218$}}
     \raise 2.0pt\hbox{$\mathchar"13E$}}}

\def\edcomment#1{\iffalse\marginpar{\raggedright\sl#1\/}\else\relax\fi}
\reversemarginpar

\begin{document}
\title{The evolution of the cluster environments of radio sources at $z<1.8$}

\author{Philip Best}
\affil{Institute for Astronomy, Royal Observatory, Blackford Hill,
Edinburgh EH9 3HJ, UK} 
\author{Matt Lehnert}
\affil{MPE, Postfach 1312, 85741 Garching-bei-Munchen, Germany}
\author{Huub R{\"o}ttgering, George Miley}
\affil{Sterrewacht Leiden, Huygens Lab, Postbus 9513, 2300RA Leiden, NL}

\begin{abstract}
An analysis of the cluster environments around distant radio galaxies is
presented, in particular the results from new NTT deep optical--IR imaging
of the fields of radio sources at $z \sim 1.6$. A net overdensity of
K--band galaxies is found, together with a sharp peak in the angular
cross--correlation amplitude, centred on the radio galaxies. This excess
clustering is associated predominantly with red galaxies, with colours
consistent with being old ellipticals at the radio source redshift. A
large excess of such red galaxies is seen, particularly within 100\,kpc of
the radio source. These comprise amongst the most distant red sequences of
cluster ellipticals yet discovered, that is, the highest redshift `normal'
clusters.\vspace*{-3mm}
\end{abstract}

\section{Introduction}

The discovery and study of clusters at redshifts beyond one is of great
importance for many aspects of structure formation and cosmology, as
discussed throughout this volume. Unfortunately, identification of
clusters at these redshifts is very challenging observationally: at X-ray
wavelengths existing surveys suffer sensitivity limits and the field of
view of the current generation of X-ray telescopes make them inefficient
for wide-area surveys; at optical and near--infrared wavelengths distant
clusters show only low contrast above the high background counts at faint
magnitudes. Fortunately an alternative method of locating distant clusters
does exist, using powerful radio sources as probes.

Radio sources are hosted by giant elliptical galaxies (e.g. McLure \&
Dunlop 2000, Best et al 1998) and are amongst the most massive galaxies
known in the early Universe. At low redshifts they are typically found in
galaxy groups, or poor clusters, but environmental richness appears to
increase with redshift: at $z \sim 0.5$, the large amplitude of the galaxy
cross--correlation function around radio galaxies and an Abell clustering
classification indicate that about 40\% of radio sources are located in
clusters of Abell richness class 0 or greater (e.g. Hill \& Lilly
1991). At $z \sim 1$ there is overwhelming evidence from observations at a
wide variety of wavelengths that at least some powerful radio sources are
located at the centres of clusters (e.g. Best 2000 and references
therein). Best (2000) analysed UKIRT K--band data of a sample of 3CR radio
galaxies at $z \sim 1$ and found a sharp peak in the cross--correlation
function surrounding the radio galaxies, corresponding to a mean
environmental richness of between Abell class 0 and 1. Colour--magnitude
(C-M) relations showed red sequences around some (but not all) radio
sources (see also Barr et al, this volume), but a detailed analysis of the
C-M relations and the source--to--source variations was prohibited by the
small field of view (70 by 70 arcsec) of the UKIRT frames.

At still higher redshifts, strong evidence is found for (proto-)clusters
around radio galaxies at $z > 2$ (e.g. Pentericci et al, Kurk et al, this
volume). At these redshifts, however, studies have been limited to objects
selected by line emission: the sequence of red galaxies characteristic of
low redshift clusters is not seen.

In this article the first results are presented of a very deep optical--IR
imaging study of the environments of radio sources at $1.44 < z < 1.7$, in
order to investigate their environments, to determine to what redshift the
centres of clusters are dominated by a population of red galaxies, and
ultimately to use this red sequence to investigate the evolution of the
cluster ellipticals.\vspace*{-3mm}

\section{The fields of radio galaxies at $z \sim 1.6$}

We have defined a complete subsample of 9 radio sources with redshifts
$1.44 < z < 1.7$ and galactic latitudes $|b|> 20^{\circ}$ from the
equatorial sample of Best et al (1999; $S_{\rm 408 MHz} > 5$\,Jy; $-30 <
\delta < +10^{\circ}$). Deep 5 by 5 arcminute R, J, K images of the fields
of five of these sources were obtained in Sept 2000 and Aug 2001 using the
NTT, reaching typical depths of $K \sim 20.5$, $J \sim 22.3$ and $R \sim
26$. Although final results await the completion of the imaging program,
these multi--colour data provide some promising early results, including:
\smallskip

\noindent $\bullet$ {\it An excess of K--band counts.} The number counts
in the infrared K--band have been determined as a function of magnitude in
each field (see Table~1). These results show a clear excess of number
counts fainter than $K \sim 18.5$. Note that the magnitude of the radio
source host galaxy is typically $K \sim 18$.
\smallskip

\begin{table}
\begin{center}
\begin{small}
\begin{tabular}{ccccccc}
$K$&Raw &Corrected & Counts per mag. & Error$^*$ & Literature & Excess \\
   &Counts & Counts & per sq. degree$^*$ &       &  Counts$^*$& Galaxies \\
15.0--15.5 &   3 &   3.0 &  308 &  177 &   218& 0\\
15.5--16.0 &   5 &   5.0 &  513 &  229 &   620& 0 \\
16.0--16.5 &   7 &   7.0 &  719 &  271 &   960& $-$1 \\
16.5--17.0 &  23 &  23.0 & 2362 &  492 &  1530& 3 \\
17.0--17.5 &  32 &  32.3 & 3320 &  587 &  2530& 3 \\
17.5--18.0 &  52 &  52.5 & 5396 &  751 &  5540& 0 \\
18.0--18.5 &  95 &  96.9 & 9958 & 1041 &  9470& 2 \\
18.5--19.0 & 128 & 130.6 & 13418& 1215 & 10500& 9 \\
19.0--19.5 & 208 & 214.4 & 22029& 1613 & 11600& 34 \\
19.5--20.0 & 276 & 293.6 & 30164& 2068 & 17200& 42 \\
20.0--20.5 & 294 & 381.8 & 39226& 3565 & 22700& 54 \\
\end{tabular}
\end{small}
\end{center}
\caption{The K--band number counts combined across the radio galaxy
fields. ($^*$: counts per magnitude per square degree.)}
\end{table}

\begin{figure}
\centerline{
\psfig{file=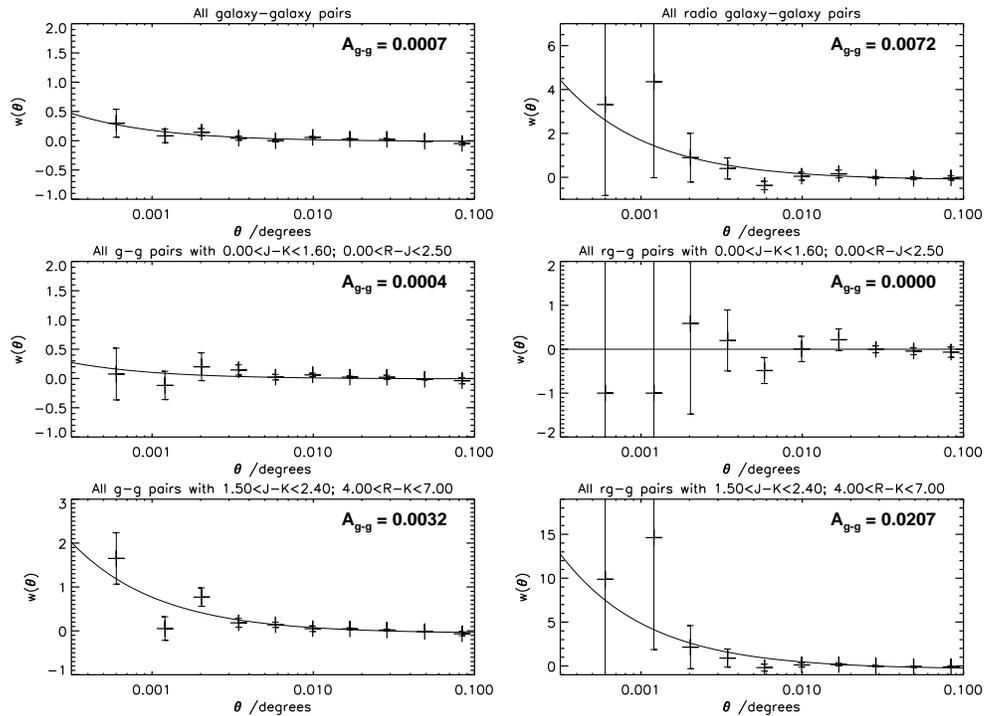,angle=90,width=13cm,clip=}
}
\caption{Angular cross--correlation analyses of the fields of the 3 radio
galaxies observed in Sept 2000. The left--hand plots consider all
galaxy-galaxy pairs, and the right--hand plots only radio galaxy--galaxy
pairs. The upper plots are for all of the galaxies in the
field, and show a clear excess of clustering around the radio galaxy. The
middle plots show blue galaxies, demonstrating that these have no
preferential clustering around the radio galaxy. The bottom plots are for
galaxies with colours consistent with being old elliptical galaxies at the
radio source redshift: these have a remarkably high cross--correlation
amplitude.}
\end{figure}

\noindent $\bullet$ {\it A peak in the angular cross--correlation function
around the radio sources.} For each field the angular cross--correlation
function has been determined following the method outlined by Best (2000);
cross--correlation statistics have been calculated both for all
galaxy--galaxy pairs, and for only radio galaxy--galaxy pairs. Combining
the data from the different fields, there is a clear excess of clustering
around the radio host galaxies, as compared to the field in general
(Figure~1), the mean cross--correlation amplitude indicating cluster
environments of Abell richness class 0 to 1. To investigate how the
cross--correlation amplitude varies with galaxy colour, the data were
split into different colour bins. The clustering excess is predominantly
associated with red galaxies; those galaxies with colours consistent with
being cluster ellipticals are very strongly clustered around the radio
galaxy, whilst blue galaxies show no excess clustering. There is some
field-to-field variation; this will be investigated when the
dataset is completed.
\smallskip

\noindent $\bullet$ {\it Overdensities of red galaxies.} In many of the
fields there is a large overdensity of extremely red galaxies (see also
Hall et al 2001), with colours consistent with passively evolving
elliptical galaxies at the radio source redshift. There is especially
strong clumping of red galaxies on a $\sim 100$kpc scale. This can be
seen, for example, in colour cuts as a function of radius (Figure
2).\vspace*{-3mm}

\begin{figure}
\centerline{
\psfig{file=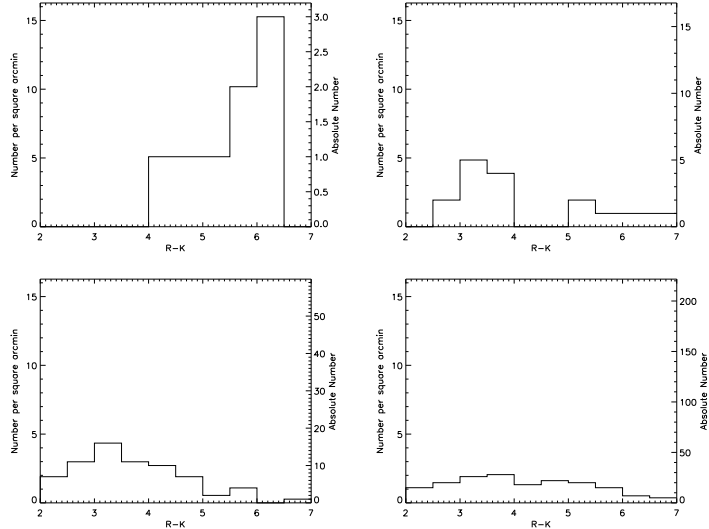,angle=0,width=9.4cm,clip=}
}
\caption{$R-K$ colour histograms for galaxies around 2025-155, as a
function of distance from the radio source: Upper left -- 0 to 100\,kpc
radius; Upper right -- 100 to 250\,kpc radius; Lower left -- 250 to
500\,kpc radius; Lower right -- 500 to 1000\,kpc radius. Note the strong
excess of galaxies with very red colours, particularly in the inner
100\,kpc.}
\end{figure}

\section{Conclusions}

We have presented some of the first results of a program of optical--IR
imaging of the fields of radio galaxies at redshifts $z \sim 1.6$. Three
different analyses all provide strong evidence for significant galaxy
overdensities around some of these radio sources. Work to
spectroscopically confirm these redshift and to study the properties of
the cluster galaxies is continuing. The presence of the extremely red
galaxies is particularly exciting, as it indicates that a substantial
population of elliptical galaxies have already formed, and must have
ceased star formation a significant time ($\gta 1$\,Gyr) before redshift
1.5. This red sequence also makes these amongst the most distant `normal'
clusters known, and thus offers unique prospects for advancing studies of
galaxy formation and evolution.
\end{document}